\documentclass[prb,twocolumn,showpacs,preprintnumbers,amsmath,amssymb]{revtex4}

\usepackage{amsfonts}
\usepackage{latexsym}
\usepackage{graphicx}
\usepackage{epsfig}
\usepackage{amsbsy}
\usepackage{amsmath}
\usepackage{amssymb}

\newcommand{\be}{\begin{equation}}
\newcommand{\ee}{\end{equation}}

\newcommand{\bear}{\begin{eqnarray}}
\newcommand{\ear}{\end{eqnarray}}

\begin{document}

\title{Corbino experimental set-up for Cooper pair
mass spectroscopy and determination of mobility in normal phase}
%Simple model of the origin of coronal temperature jump

\author{T.M.~Mishonov}
\email[E-mail: ]{mishonov@phys.uni-sofia.bg}
\author{M.V.~Stoev}
\email[E-mail: ]{martin.stoev@gmail.com}

\affiliation{Department of Theoretical Physics, Faculty of Physics,\\
University of Sofia St.~Clement of Ohrid,\\
5 J. Bourchier Boulevard, Bg-1164 Sofia, Bulgaria}

\date{\today}

\begin{abstract}
We are suggesting an electronic method for Cooper pair mass
spectroscopy of thin superconducting films. The method can be
applied, for example, for 100~nm thin high-$T_c$ films grown on
insulator substrate 10~mm$\times$10~mm. In Corbino geometry two
Ohmic contacts have to be made on the film evaporating Ag or using
silver paste: one circle with radius $r_a$ (1) and a ring electrode
with internal radius $r_b$ (2). In the in-between space
($r_a$,$r_b$) a ring electrode from mylar assures a capacitive
connection between the superconducting layer and the metallized
surface (3) of the mylar. In such a way we have an field effect
transistor (FET) type structure with a circular gate. When at low
enough temperatures $T\ll T_c$ an AC current with frequency $\omega$
is applied between the circle source (1) and the ring-shaped drain
(2) an AC Bernoulli voltage with double frequency $2f$ appears
between the gate (3) and the source (1). The $2f$ signal depends on
Cooper pair effective mass and its systematic investigation gives a
Cooper pair mass spectroscopy. In the normal phase $2f$ gives
logarithmic derivative of the density of states (DOS) with respect
of Fermi energy. Applying a gate voltage in the same structure gives
the mobility of the normal phase.
\end{abstract}


\maketitle

%%%%%%%%%%%%%%%%%%%%%%%%%%%%%%%%%%%%%%%%%%%%%%%%%%%%%%%%%%%%%%%%%%%%

In spite of the intensive development of physics of
superconductivity and especially high-$T_c$ superconductivity in
during the last two decades the effective mass of Cooper pairs
remains an almost inaccessible parameter.\cite{Mishonov:00} The
purpose of the preset work is to suggest the simplest electronic
method for determination of the effective mass which does not
require patterning of the film or preparing of superconducting
micro-structures. The method can be applied for bare superconducting
films; imagine for example a 100~nm cuprate superconductor grown on
a 10~mm$\times$10~mm perovskite substrate. The method we suggest is
based on the Bernoulli effect.

The main detail of the suggested experimental set-up is a ring
electrode (3) from mylar affixed on the clean superconductor
surface. The polymer insulator layer of the mylar contacts to the
superconducting surface while the gate electrode is affixed on the
top of the metallized layer of the mylar. In  such a way we have a
metal-insulator-superconductor (MIS) type structure. The ring has
inner radius $r_a$ and outer radius $r_b$. Using a silver paste or
evaporating silver, a circular Ohmic contact (1) with radius $r_a$
is made in the center. Another Ohmic probe (2) with inner radius
$r_b$ surrounds the capacitor ring. The Ohmic contacts (1) and (2)
are as source and drain probes of a field-effect transistor (FET)
type structure, while the metallized layer (3) plays the role of the
gate. Thus we have a superconductor FET-type device in circular
Corbino geometry.

When a source-drain current
\be I_{SD}(t)=I_0\cos{\omega t} \label{Current(1f)} \ee
is applied between source (1) and drain
(2) an almost homogeneous current density is created
\be
j=\frac{I_{SD}}{2\pi r d_{\mathrm{film}}}=e^{*}nv
\label{Density}
\ee
in the superconductor layer. We suppose that the thickness of the
film is smaller than London penetration depth $\lambda$. For
simplicity we consider low enough temperatures $T\ll T_c$ when all
charge carriers are superfluid and the penetration depth is
expressed by the effective mass of Cooper pairs $m^*$, its charge
$e^*$ and volume density $n$
\be
\frac{1}{\lambda^2(T)}=\frac{e^{*2}n}{m^*c^2\epsilon_0}; \quad
\frac{1}{c^2\epsilon_0}=\mu_0=4\pi*10^{-7}.
\label{Penetration}
\ee
For dissipation free low frequency current, the constancy of the
electrochemical potential in the bulk of the superconductor for
$T\ll T_c$ gives Bernoulli theorem
\be
\frac{1}{2}m^*v^2+e^*\phi(r)=const.
\label{Bernoulli}
\ee
The radius dependent electric potential $\phi(r)$ creates the
Bernoulli voltage
\be
U_b(t)=-\frac{1}{2}\frac{m^*}{e^*}\langle v^2\rangle.
\label{Ub}
\ee
Where averaging is on the superconducting surface of the plane
capacitor.

For this averaging we have to calculate
\be
\langle\frac{1}{r^2}\rangle=
\frac{\int_{r_a}^{r_b}{\frac{1}{r^2}\mathrm{d}\pi r^2}}
{\int_{r_a}^{r_b}{1\, \mathrm{d}\pi r^2}}=
\frac{\ln(r_b^2/r_a^2)}{r_b^2-r_a^2}.
\label{Averaging}
\ee
And for Bernoulli voltage measured at second harmonics
\be
U(t)=U_{2f}\cos{2\omega t}
\label{U(t)}
\ee
we obtain
\be U_{2f}=-\frac{\ln(r_b/r_a)}{4\pi(r_b^2-r_a^2)d_{\mathrm{film}}}
\frac{\lambda^2(0)}{\lambda^2(T)}
\left(\frac{\lambda^2(0)}{\epsilon_0 c^2}\right)^2
\frac{e^*}{m^*}\,I_0^2.
\label{U2f} \ee
The plot $U_{2f}$ versus $I_0^2$ determines $m^*$. The temperature
dependance $\lambda^2(0)/\lambda^2(T)$ is described in
Refs.~\onlinecite{Mishonov:00,Mishonov:94}. The Bernoulli potential
as current induced contact potential difference \cite{Mishonov:94}
has to be measured as source~(1) -- gate~(3) signal. Additional
possibilities are: for a second point to be used not the central
ring but some Ohmic connection to certain of the corners of the
films where the current density is very low or a second capacitor
ring (4) to be affixed surrounding the other Ohmic ring. If the
capacitance of the mylar ring is not much smaller than the internal
capacitance of the Lock-in preamplifier we have to take into account
the corresponding decrease of the signal. The evaluations of the
magnitude for cuprate films gives that Bernoulli potential can be of
order of $\mu\mathrm{V}$ and the experiment is actually very simple.

Few words we have to add for the normal phase. Above $T_c$ the
normal current oscillations in this case create the oscillations of
the temperature with frequency $2f$. Second harmonic oscillations of
the work function $W(t)$ give the logarithmic derivative of the
density of states.\cite{Mishonov:06}

In such a way the suggested experimental set-up can give important
fundamental knowledge for the properties of the investigated
materials. It is a pity that up to now we have only one reliable
determination of effective mass of Cooper pairs based actually on a
similar MIS structure investigations.\cite{Mishonov:91,Fiory:90} If
silver paste is used for making of Ohmic contact they can be often
cleaned by acetone and the film can be reused for other studies.

Every Lock-in has sensitivity of $\mu\mathrm{V}$ and possibility of
second harmonic detection. That is why we conclude that the
suggested method for mass spectroscopy can be realized in every
laboratory involved with preparation of high-$T_c$ films, and in
every low temperature physics laboratory effective mass of Cooper
pairs $m^*$ can be measured. If the gate electrodes are used not for
the detection but to apply an external voltage, a lot of new
possibilities for fundamental research are open. The gate voltage
induces external charge density on the surface of the superconductor
\be
D_z=\epsilon_0\epsilon_r E_z;\quad
E_z=U_{\mathrm{gate}}/d_{\mathrm{ins}}.
\label{Charge_density}
\ee
As it is typical for physics of FET these surface charges change the
two dimensional conductivity which in Corbino geometry is radial
\be
j_r^{(2D)}=(en_{\mathrm{normal}}d_{\mathrm{film}}+D_z)\frac{e\tau}{m}E_r
\label{Current_density}
\ee
Investigation of the SD current $j_r^{(2D)}$ influenced by the gate
voltage can give an independent method for determination of mobility
$e\tau/m$ of the normal phase and charge carrier density.

Another possibility to determination of effective mass of Cooper
pairs is to use electrostatic doping of the superconducting phase
and its influence on the eddy currents created by a DC coil. Perhaps
it is a more convenient procedure than the original method applied
by Fiory and al.\cite{Fiory:90}

\acknowledgments

Support and fruitful discussions with D.~Damianov are highly
appreciated.

%%%%%%%%%%%%%%%%%%%%%%%%%%%%%%%%%%%%%%%%%%%%%%%%%%%%%%%%%%%%%%%%%%%%

\end{document}